\documentclass[9pt,twocolumn,twoside]{osajnl}
\journal{ol} % Choose journal (ao,jocn,josaa,josab,ol,optica,pr)

\setboolean{shortarticle}{true}
\usepackage{lineno}
\title{Mid-infrared edge-enhanced imaging via angle-selective nonlinear filtering}

\author[1]{Zhuohang Wei}
\author[1,2,3,*]{Kun Huang}
\author[1]{Jianan Fang}
\author[1,2,4]{Heping Zeng}

\affil[1]{State Key Laboratory of Precision Spectroscopy, East China Normal University, Shanghai 200062, China}
\affil[2]{Chongqing Key Laboratory of Precision Optics, Chongqing Institute of East China Normal University, Chongqing, China}
\affil[3]{Collaborative Innovation Center of Extreme Optics, Shanxi University, Taiyuan, Shanxi 030006, China}
\affil[4]{Shanghai Research Center for Quantum Sciences, Shanghai 201315, China}
\affil[*]{khuang@lps.ecnu.edu.cn}

\begin{abstract}
Mid-infrared reconfigurable edge-enhanced imaging is highly demanded in sensing and vision fields. Here, we propose a novel scheme for mid-infrared upconversion imaging with high tunability between bright-field and edge-enhanced modalities. The involved engineering of the nonlinear process favors shaping the optical transfer function of the imaging system. Consequently, a nonlinear angle-selective filter can be configured to perform an all-optical Fourier processing of the image, which highly depends on phase-matching parameters. We numerically demonstrate the ability to switch modalities between the bright-field and edge-enhanced imaging by tuning the crystal temperature, and to simultaneously acquire both information by dichromatic illumination. Notably, the achieved reconfigurability is realized without changing the imaging settings, which contrasts to previous instantiations based on pump adaptation. Therefore, the proposed architecture of upconversion imagers would pave a novel way to implement layout-compact and all-optical processing for infrared images.
\end{abstract}

\setboolean{displaycopyright}{true}

\begin{document}

\maketitle
Edge-enhanced detection, as a critical technique in pattern recognition and machine vision, constitutes a powerful tool for image analysis in a wide range of fields, such as bio-tissue imaging, atmospheric observation, and material inspection \cite{Zhou2020NP}. Nowadays, the edge enhancement of recorded intensity images is typically performed by post-processed computation based on differentiation operation or Hilbert transformation \cite{He2022NPh}. Compared with the traditional digital method, all-optical analog processors offer the advantages of low energy consumption, short latency time, and fast processing speed \cite{Zangeneh2021NRM}. Common all-optical edge detection relies on Fourier filtering in a 4f configuration, where Fourier components are selectively filtered by a spatially varying amplitude and/or a phase mask placed in the Fourier plane \cite{Davis2000OL, Huo2020Nanoletter}. Alternatively, one can engineer the Green's function (\textit{i.e.}, the impulse response) of the optical system, which has facilitated promising optical computing systems based on meta-optical and thin film devices \cite{Wesemann2021APR}. The resultant angle-sensitive filters perform image Fourier processing directly on the object plane \cite{Guo2018Optica}, thus favoring compact and tunable edge detection \cite{Cotrufo2023NC, Cotrufo2024NC}. However, the reported all-optical technologies for tunable edge enhancement are mostly restricted in the visible and near-infrared bands. Hence, it becomes of significant importance to extend the operational window into longer-wavelength spectral regions, with an aim to promote expanding applications.

Particularly, mid-infrared (mid-IR) imaging provides desirable features in detecting thermal radiation, analyzing chemical compositions, and penetrating environmental obscurants, which finds widespread applications including biomedical examination, night vision, and military surveillance \cite{Rogalski2011IPT}. However, existing mid-infrared detectors suffer from pressing difficulties in high dark noise, limited pixel format, and  low frame rate, which severely limit the imaging performance in terms of sensitivity, resolution, and speed \cite{Wang2019Small}. Additionally, emerging micro- and nano-materials based on silica substrates have high absorption in the mid-IR regime. Fabrication technologies compatible with efficient and tunable manipulation of infrared optical fields are still under development \cite{Hu2024AP}. It has been a long-standing quest to develop a high-efficiency, low-noise, and tunable mid-IR edge enhancement imaging technology.

In this context, the frequency upconversion strategy provides an indirect approach for mid-IR sensing, where infrared information is nonlinearly converted to the visible or near-infrared spectrum, thereby allowing one to fully leverage silicon detectors to achieve sensitive and fast mid-IR detection and imaging \cite{Barh2019AOP}. Moreover, the manipulation of the pump field can be used to engineer the imaging functionality through nonlinear wave mixing \cite{Qiu2018Optica}. Specifically, upconversion edge-enhanced imaging has been implemented by applying a vortex phase to the pump, which facilitates simultaneous operations of high-fidelity optical modulation and high-sensitivity upconversion detection \cite{Liu2019PRA}. In this scenario, the crystal is typically placed at the Fourier plane in a 4f configuration, which requires careful preparation of a complex pump field to control spatial frequency components at the Fourier plane \cite{Wang2021LPR, Li2024OLT}. Indeed, the pump property basically determines the system's transfer function, while the change of phase-matching parameters or illumination conditions merely alters the field of view \cite{Junaid2020AO}. Therefore, the 4f-based upconversion imaging configuration usually results in a single and static imaging modality as predefined by the system settings, which imposes restriction in applications where modality switching between bright-field and edge-enhanced imaging, or parallel acquisition of both information is needed \cite{Zhu2024NL, Badloe2023ACSNano, Zhang2021NL}.

In this work, we propose a novel scheme for reconfigurable mid-IR edge-enhanced upconversion imaging, which excludes the requirements for sophisticated tailoring of the pump field and precise alignment along the optical axis in the conventional 4f configuration. Here, the nonlinear crystal is placed close to the object to directly address the angular spectrum. The system's transfer function is sensitive to incident wavelength and crystal temperature, thus enabling angle-selective upconversion for the spatial frequency components of the object. This presented paradigm offers advantages such as compact imaging layout, reconfigurable transfer function, and the capability to simultaneously acquire bright- and dark-field information.

%%%%%%%%%%%%%%%%%%%%%%%%%%%%%%%%%%%%%%%%%%%%%%%%%%%%%%%%%%%%

\begin{figure}[b!]% or add * to become figure*
\centering
\includegraphics[width=0.4\textwidth]{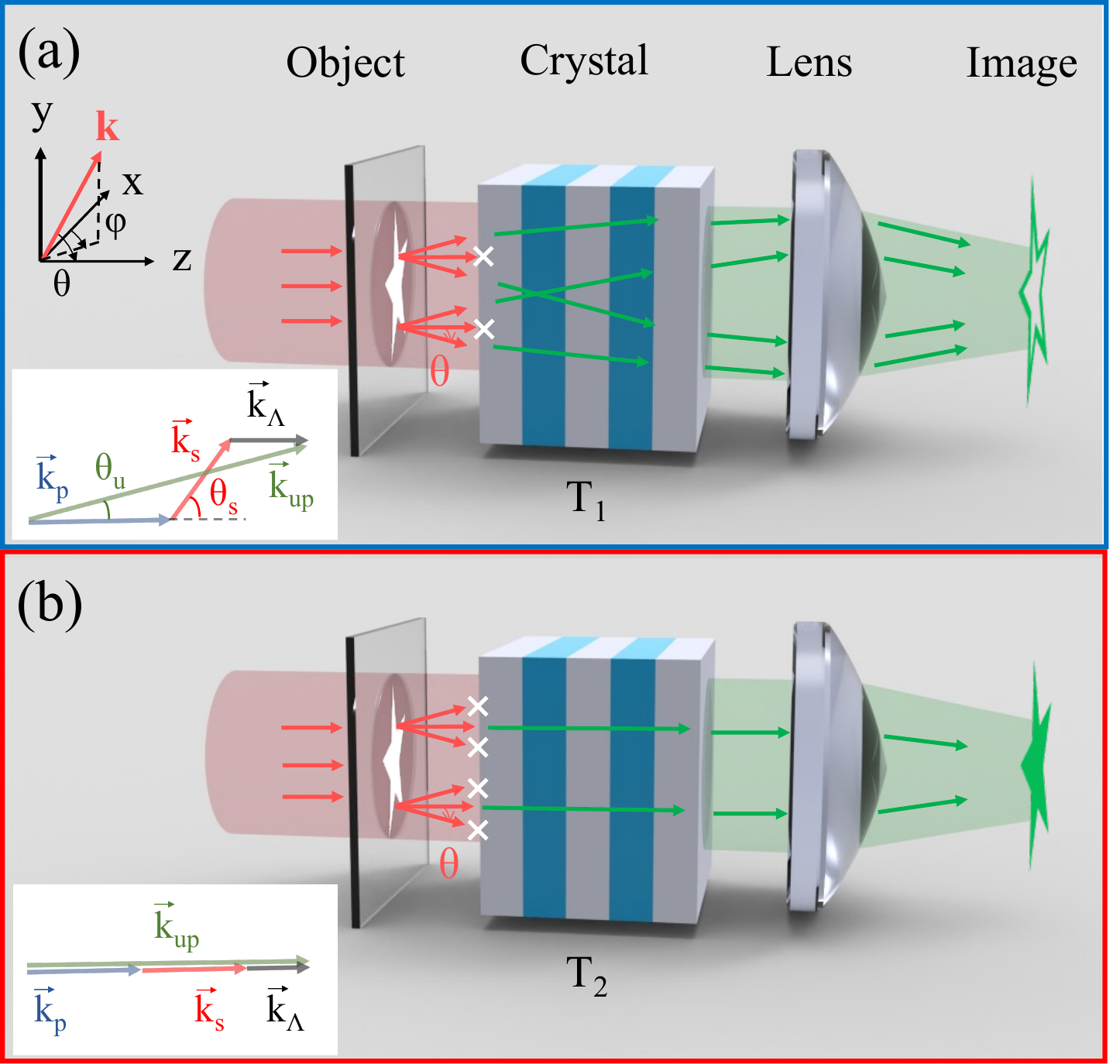}
\caption{Scheme of angle-selective upconversion imaging at bright- and dark-field modalities. (a) Dark-field imaging. (b) Bright-field imaging. Insets at the left-bottom corner show the phase-matching diagrams for the involved wave vectors.}
\label{fig1}
\end{figure}

We start with an introduction to the all-optical Fourier processing. Under coherent illumination, an object can be decomposed into a bundle of plane waves at various propagating directions, as shown in Fig. \ref{fig1}. The transverse spatial frequencies can be expressed as $[k_x, k_y] = k_0\sin\theta[\cos\phi, \sin\phi]$, where $\theta$ and $\phi$ are the polar and azimuthal angles, ${{k}_{0}}={2\pi }/{\lambda }$ is the wavenumber, $\lambda$ denotes the wavelength. For an object with an electric field of $E_\text{in}(x,y)$, the spatial frequency distribution is proportional to the Fourier transform $f_\text{in}(k_x,k_y)=\iint E_\text{in}(x,y)e^{-i(k_xx+k_yy)} dx dy$. After an imaging system with a transfer function of $H(k_x,k_y)$, these plane waves are collected and re-creating the output image $E_\text{out}(x,y)$ at a different plane, effectively performing an inverse Fourier transform as
\begin{equation}
E_\text{out}(x,y) = \iint H(k_x,k_y) f_\text{in}(k_x,k_y) e^{i(k_xx+k_yy)} dk_x dk_y \ .
\label{Eq1}	
\end{equation}
In principle, any operation in the Fourier space can be performed by selectively filtering the bundle of plane waves originating from the image with respect to their propagation direction \cite{Cotrufo2023NC}. As shown in Fig. \ref{fig1}(a), we can realize a high-pass filter by suppressing plane waves at small propagating angles ($\theta \approx 0$), while transmitting waves propagating at larger angles. The resultant momentum filtering allows us to enhance the edges of an input image with respect to homogeneous regions in Fig. \ref{fig1}(b). Such an angle-selective filtering can be implemented with a suitably designed nonlinear optical process in our scheme. By tailoring the phase-matching conditions of the nonlinear crystal, the desired mathematical operation can be encoded into the optical transfer function of the upconversion imaging system.

Specifically, the involved frequency upconversion occurs in a periodically poled lithium niobate (PPLN) crystal based on sum-frequency generation (SFG). The energy conversion requires $\hbar {{\omega }_{u}}=\hbar {{\omega }_{s}}+\hbar {{\omega }_{p}}, $ where $\hbar$ is the Planck constant, ${{\omega }_{s,p,u}}$ are the angular frequencies of signal, pump, and SFG fields, respectively. In addition, the momentum conservation should be satisfied to achieve the quasi-phase-matching condition \cite{Huang2022NC}. As illustrated in Fig. \ref{fig1}, the axial mismatch of wave vectors is given by
\begin{equation}
	\Delta k =2\pi \left[ \frac{{{n}_{u}}}{{{\lambda }_{u}}}\cos \left( \frac{{{\lambda }_{u}}}{{{\lambda }_{s}}}{{\theta }_{s}} \right)-\frac{{{n}_{s}}}{{{\lambda }_{s}}}\cos {{\theta }_{s}}-\frac{{{n}_{p}}}{{{\lambda }_{p}}}-\frac{1}{\Lambda } \right] \ ,
	\label{Eq2}	
\end{equation}
where ${{n}_{u,s,p}}$ denote the refractive indices, ${{\theta }_{s}}$ is the incident angle of signal beam, and $\Lambda $ is the poling period of nonlinear crystal. Here, an approximation is used ${{{\theta }_{u}}}/{{{\theta }_{s}}} \simeq {{{\lambda }_{u}}}/{{{\lambda }_{s}}}$, which stems from the transverse phase matching. Under undepleted pump approximation, the SFG intensity is given as $I\left(T,{{\lambda }_{s}},{{\lambda }_{p}},{{\theta }_{s}} \right)\propto \text{sinc}^2\left[ {\Delta k\left(T,{{\lambda }_{s}},{{\lambda }_{p}},{{\theta }_{s}} \right)L}/{2}\right]$ after propagating through the crystal with a length $L$ at a temperature $T$. This formula indicates that efficient conversion only takes place at $\Delta k \to 0$. By properly adjusting the phase-matching parameters of $T$ and $\lambda_s$, plane-wave components within a specific range of incident angles $\theta_s$ can be selectively upconverted. To this end, we place the crystal close to the object to perform the nonlinear angle-selective operation, which maximizes the amount of scattered light captured directly through the nonlinear filter. In the paraxial approximation, the transfer function is written as:
\begin{equation}
	\begin{aligned}
		& H({{k}_{x}},{{k}_{y}})\propto \text{sinc}\{\pi L[\frac{{{n}_{u}}}{{{\lambda }_{u}}}\cos (\frac{{{\lambda }_{u}}\sqrt{k_{x}^{2}+k_{y}^{2}}}{{{k}_{0}}{{\lambda }_{s}}}) \\
		&-\frac{{{n}_{s}}}{{{\lambda }_{s}}}\cos (\frac{\sqrt{k_{x}^{2}+k_{y}^{2}}}{{{k}_{0}}})-\frac{{{n}_{p}}}{{{\lambda }_{p}}}-\frac{1}{\Lambda }]\} \ .
	\end{aligned}
	\label{Eq3}	
\end{equation}
For given parameter settings, the upconversion image can be evaluated by combining Eq. \ref{Eq1} and Eq. \ref{Eq3}. 

Furthermore, a numerical model is established to investigate the evolving two-dimensional field distribution along the interaction length. Rigorously, the upconversion process can be described by the coupled wave equation that deals with the coupling of waves of different frequencies in the nonlinear parametric interaction \cite{Liu2019PRA}:
\begin{equation}
	i2{{k}_{u}}\frac{\partial {{E}_{u}}}{\partial z}+\nabla _{\bot }^{2}{{E}_{u}}=-\frac{\omega _{u}^{2}{{d}_\text{eff}}}{{{\varepsilon }_{0}}{{c}^{2}}}{{E}_{s}}{{E}_{p}}{{e}^{i\Delta kz}} \ ,
	\label{Eq4}	
\end{equation}
where ${{E}_{u,s,p}}$ denote the complex amplitudes of the three involved fields, $z$ is the propagation distance within the crystal, ${{\varepsilon }_{0}}$ is the vacuum permittivity, $c$ is the speed of light in vacuum, and ${{d}_\text{eff}}$ is the effective nonlinear coefficient. The simulation procedure consists of two major parts: free-space propagation and nonlinear propagation \cite{Coen2023OE}. The free-space propagation is calculated through the Fresnel diffraction formula:
\begin{equation}
	\begin{aligned}
	    &E(x',y',d)=-\frac{\exp (-ikd)}{i\lambda d} \\
	    & \iint{E(x,y,0) 
		 \exp \left\{ \frac{i\pi }{\lambda d}\left[ {{\left( {x}'-x  \right)}^{2}}+{{\left( 
    	 {y}'-y \right)}^{2}} \right] \right\}dxdy} \ ,
	\end{aligned}
	\label{Eq5}	
\end{equation}
where $d$ denotes the propagation distance.  $[x,y]$ and $[{x}',{y}']$ represent the coordinates in the object and image planes, respectively. Once the optical fields enter the nonlinear crystal, the nonlinear interaction should be included besides the diffraction effect \cite{Liu2019PRA}. The resultant nonlinear propagation can be calculated based on the split-step Fourier approach \cite{Coen2023OE}, as given by
\begin{equation}
	\begin{aligned}
		& {{E}_{u}}\left( {x}',{y}',z+dz \right)={{E}_{u}}\left( x,y,z \right) \\
		&+\frac{i}{2{{k}_{u}}}\frac{\omega_{u}^{2}{{d}_\text{eff}}}{{{\varepsilon}_{0}}{{c}^{2}}}{{F}_\text{crystal}}(z)
	    {{\mathcal{F}}^{-1}}[\mathcal{F}({{E}_{s}})H\left( {{k}_{x}},{{k}_{y}} \right)]
	    {{E}_{p}}{{e}^{i\left( {{k}_{s}}+{{k}_{p}}-{{k}_{u}} \right)z}}dz \\ 
		& -\frac{i}{2{{k}_{u}}}{{\mathcal{F}}^{-1}}
		\left\{ \left( k_{x}^{2}+k_{y}^{2} \right)\mathcal{F}\left[ {{E}_{u}}\left( x,y,z \right) \right] \right\}dz \ ,
	\end{aligned}
	\label{Eq6}	
\end{equation}
where ${{F}_\text{crystal}}\left( z \right)=\text{sign}\left[ \cos \left( {2\pi z}/{\Lambda}\right) \right]$ is the modulation function of the crystal, $dz=L/N$ is the simulation step length, and $N$ is the simulation step number. ${{k}_{s,p,u}}$ are the wave vectors of signal, pump and upconversion beams within the crystal, respectively. For the system in Fig. \ref{fig1}, the upconversion beam passes through a lens with a phase modulation term $\exp \left[ {-i\pi \left( {{x}^{2}}+{{y}^{2}} \right)}/{\left( \lambda f \right)}\right]$ to form the image, where $f$ denotes the focus length of lens. With all these propagators for the linear and nonlinear elements, the upconversion image can be obtained through a sequence of numerical computations for cascaded sectors within the setup.

\begin{figure}[b!]
\centering
\includegraphics[width=0.95\columnwidth]{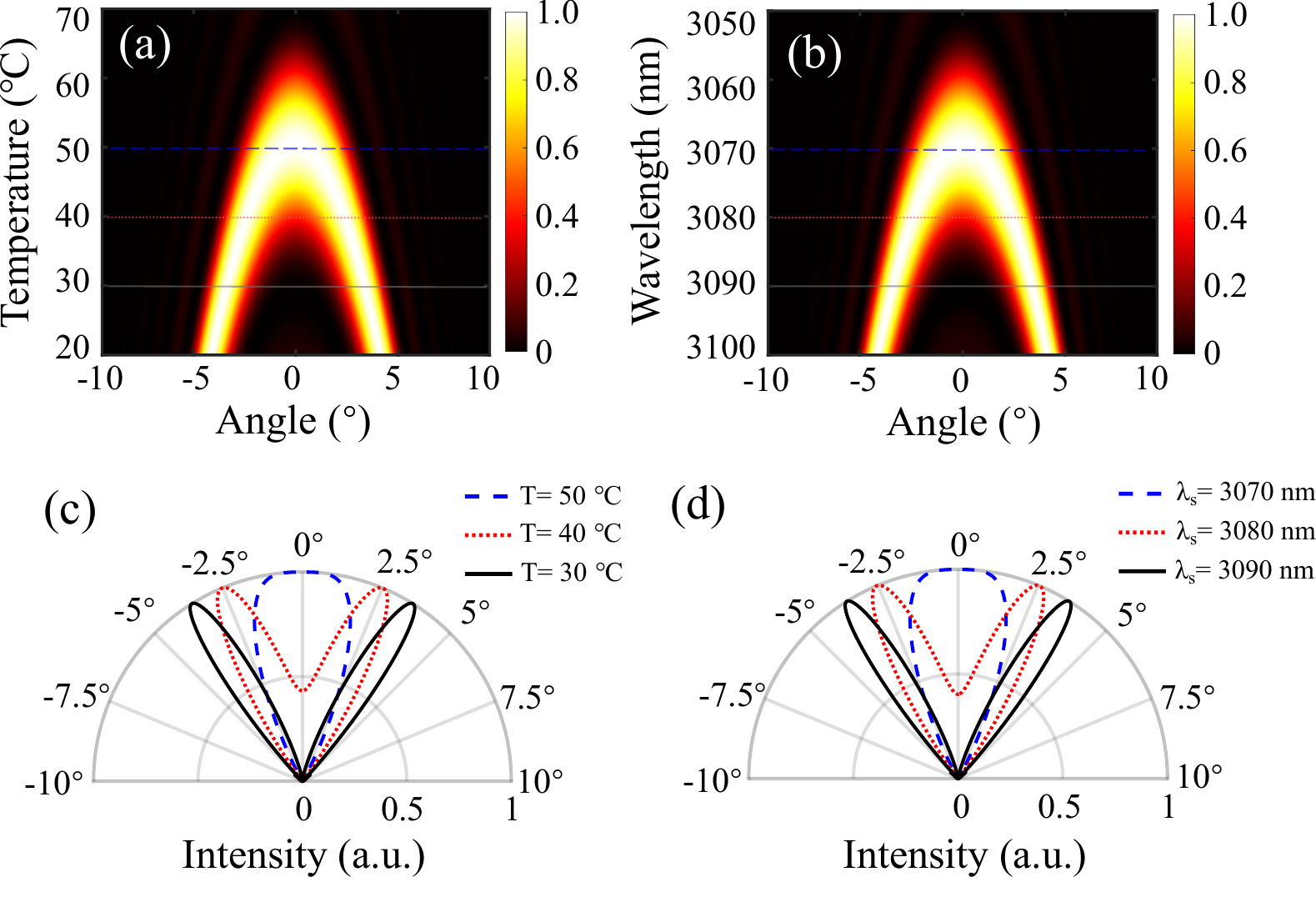}
\caption{Upconversion intensity at different incident angles. (a, b) Intensity map as a function of the crystal temperature (a) and signal wavelength (b). (c, d) Corresponding cross sections along the dashed lines denoted in the maps (a, b).}
\label{fig2}
\end{figure}
  
First, we calculate the transmission of this nonlinear filter at different incident angles. Fig. \ref{fig2} presents the normalized upconversion intensity (that can also be regarded as the normalized conversion efficiency) as varying the crystal temperature and signal wavelength. Here, the crystal length $L$ is 5 mm, and the poling period is 20.9 $\mu$m. The phase-matching temperature is 50 $^\circ$C. The signal and pump wavelengths are set to be 3070 and 1030 nm, which corresponds to an upconversion wavelength at 771 nm. Figure \ref{fig2}(a) shows the upconversion intensity for a temperature tuned from 20 to 70 $^\circ$C. The corresponding cross sections are given in Fig. \ref{fig2}(c) at temperatures of 50, 40, and 30 $^\circ$C. It can be seen that the optimum phase-matching angle deviates from the normal incidence as the crystal temperature decreases. As a result, the signal wave vectors at small incident angles will be cut off, which correspond to low spatial frequency components. Instead, the high-frequency components around the optimum angle are preferentially upconverted to form an edge-enhanced image. Similarly, Fig. \ref{fig2}(b) shows angle-dependence upconversion intensity at various signal wavelengths from 3070 to 3090 nm. The corresponding cross sections are depicted in Fig. \ref{fig2}(d) at wavelengths of 3070, 3080, and 3090 nm, respectively. In the case of longer-wavelength illumination, the low-frequency components will be blocked, thus leading to dark-field modality. The demonstrated flexible functionality of nonlinear angular filtering in the 2f-based scheme constitutes the core to manipulate the transfer functions of the optical system, which in turn leads to reconfigurable imaging performances between bright- and dark-field modalities.

\begin{figure}[b!]
\centering
\includegraphics[width=0.9\columnwidth]{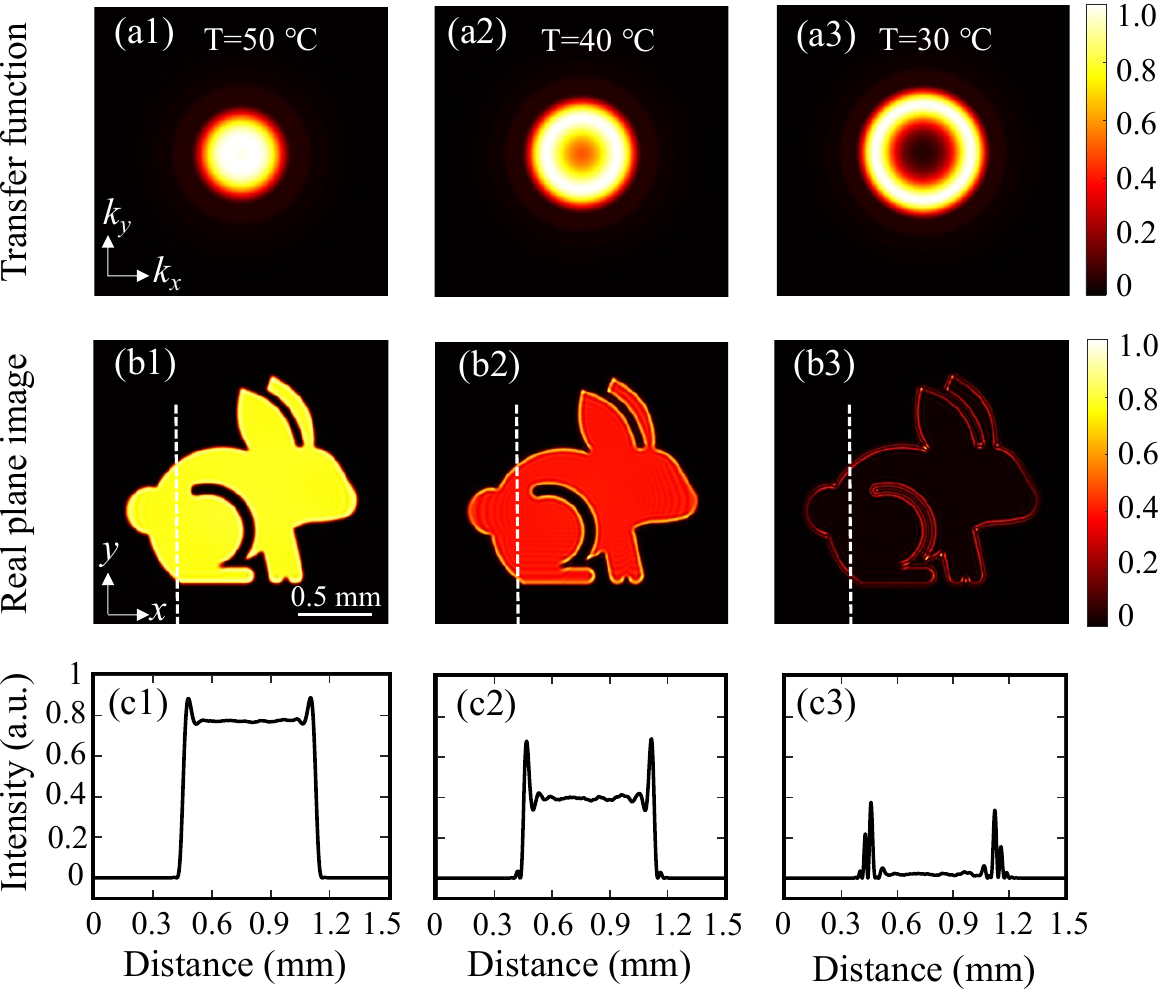}
\caption{Reconfigurable mid-IR upconversion imaging via temperature tuning. (a1-a3) Transfer function at T=50, 40, and 30 $^\circ$C. (b1-b3) Upconversion images. (c1-c3) Cross sections along the dashed lines denoted in the images (b1-b3).}
\label{fig3}
\end{figure}

Then, we characterize the imaging performance by dynamically adjusting the transfer function. The tuning operation is performed by varying the crystal temperature, without the need to change the system setup. In the simulation, the transverse space is digitized into a grid with 2048$\times$2048 points for an interrogated area of 2$\times$2 mm$^2$. These simulation settings are optimized between the calculation precision and speed. The lens has a focus length of 50 mm, and has an aperture radius of 10 mm. The distance $z_1$ between the object and the front of the crystal is set to be 1 mm, while the distance $z_2$ between the end of the crystal and the lens is 54 mm. The image is formed at $z_3=300$ mm behind the lens. The simulation step number is taken to be 250 along the nonlinear propagation length of 5 mm. In our numerical analysis, all the parameters are designated with the reference to the practical experimental settings \cite{Wang2021LPR, Huang2022NC}, such as the temperature tuning range, periodical-poling structures, and laser illumination wavelengths. Figures \ref{fig3}(a1-a3) present the transfer functions at $T$ = 50, 40 and 30 $^\circ$C at a fixed illumination wavelength of 3070 nm. At $T$ = 50 $^\circ$C, the transfer function is close to a Gaussian profile. Only components with low spatial frequency can be upconverted, thus resulting in a bright-field image as shown in Fig. \ref{fig3}(b1). As the temperature decreases, the transfer function gradually expands outward to form a donut shape, where the conversion efficiency for the low-frequency components around the center becomes diminished. In Fig. \ref{fig3}(b2), an edge sharpening effect slightly emerges at  $T$ = 40 $^\circ$C. The edge enhancement is more prominent at a smaller temperature of 30 $^\circ$C, as shown in Fig. \ref{fig3}(b3). Figures \ref{fig3}(c1-c3) give the cross sections of upconversion images along the dashed lines. In Fig. \ref{fig3}(c3), a high edge-detection contrast is obtained due to the efficient suppression of the low-spatial-frequency components of the input image. Notably, each strong edge in the input image results in two split peaks in the upconversion image, as expected from the second-order differentiation of a step-like function. The presence of additional weaker intensity fluctuations is ascribed to the ringing artifacts at sharp boundaries caused by the finite numerical aperture of the optical system \cite{Cotrufo2023NC, Cotrufo2024NC}, which is also manifested in the bright-field image. The dynamic switching process is given in \textbf{Visualization 1}, which indicates a decreased upconversion power from bright-field to edge-enhanced modalities. Since the power for a natural image mostly concentrates at the low-frequency region, the bright-field imaging typically favors a higher power conversion efficiency.

\begin{figure}[t!]
\centering
\includegraphics[width=0.7\columnwidth]{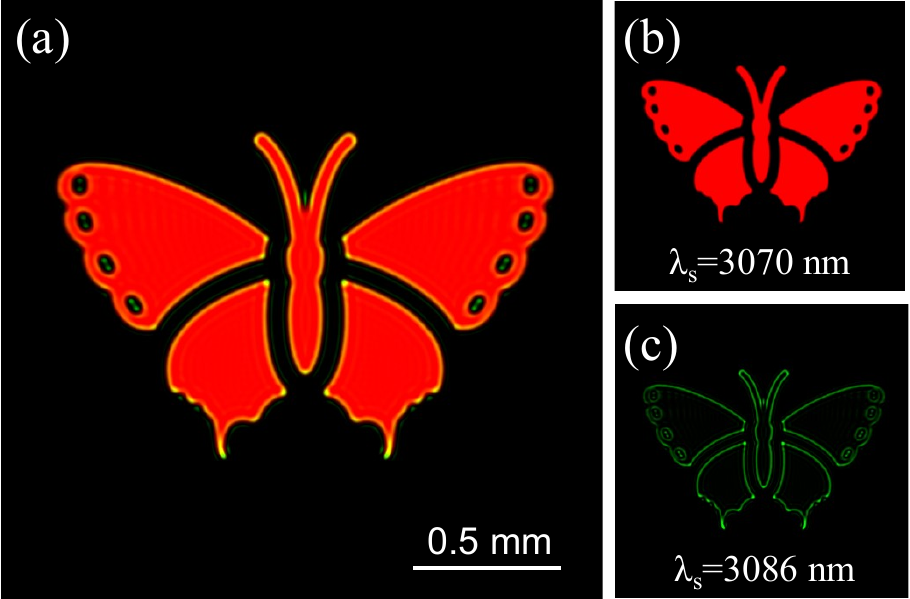}
\caption{Modality-fusion mid-IR imaging. (a) Upconversion images under dichromatic illumination. (b) Bright-field image at $\lambda_s$ = 3070 nm. (c) Dark-field image at $\lambda_s$ = 3086 nm.}
\label{fig4}
\end{figure}

Finally, we investigate the modality-fusion imaging performance under dual-wavelength illumination, where bright-field imaging and edge-enhanced imaging operate at the same time to extract different morphological information on an object. Figure \ref{fig4}(a) shows an upconversion image for an object that is coherently illuminated at both 3070 nm and 3086 nm for a fixed operation temperature of 50 $^\circ$C. Intriguingly, the upconversion image contains both the bright-field and dark-field information. At $\lambda_s$ = 3070 nm, a bright-field image is obtained in Fig. \ref{fig4}(b). At another signal wavelength of $\lambda_s$ = 3086 nm, we can capture a dark-field image as shown in Fig. \ref{fig4}(c). The dynamic transition between two imaging modalities can be implemented in an agile fashion by resorting to the illumination sources with fast-switching wavelengths based on the acousto-optic tuning technique \cite{Fang2024NC}. With the help of dispersive elements, such as an optical prism or a grating, the modality-fusion upconversion image be simultaneously captured at different locations on a silicon camera, which leads to the snapshot acquisition of the bright- and dark-field information of the targeted object. Notably, in the presented 2f-based imaging system, the field of view and the spatial magnification are independent of the signal wavelength, which eliminates the radial dispersion effect observed in conventional 4f-based upconversion imagers \cite{Fang2024NC}. Therefore, the upconversion images at various signal wavelengths can be overlapped spatially, which avoids additional data processing for imaging correction. This wavelength-independent spatial scaling is manifested by the recorded sequence of upconversion images under a broadband illumination from 3070 to 3094 nm with a step of 0.5 nm, as shown in \textbf{Visualization 2}.

In summary, we propose a novel scheme for mid-IR upconversion imaging with a unique ability to dynamically reconfigure Fourier processing operations. The all-optical image processing is implemented without requiring to physically access the Fourier space, which contrasts to previous upconversion imagers based on 4f configuration. Particularly, the approach uses angle-selective nonlinear filtering to directly satisfy the desired transfer function in the wave vector domain, thus resulting in a more compact system that does not require alignment of the filter along the optical axis. By adjusting the crystal temperature, the bright-field and edge-enhanced imaging modalities can be switched in the same setup. Moreover, modality-fusion upconversion images can be recorded with a dual-color illumination. Notably, the operation wavelength window is limited to about 5 $\mu$m for the PPLN crystals, which could be extended to far-infrared or terahertz regions by using AgGaS$_2$ or GaP crystals \cite{Barh2019AOP}. Additionally, the required phase-matching condition is fulfilled at the single-polarization operation. The processing compatibility for arbitrary polarizations is possible by resorting to two cascaded nonlinear crystals with orthogonal positions. Therefore, such the nonlinear momentum filtering operation not only enriches the capability for infrared upconversion imaging systems, but also opens up new possibilities to manipulate the transfer functions for accessing advanced computing functionalities in image processing \cite{Cotrufo2023NC, Cotrufo2024NC}  or wavefront sensing \cite{Liu2023LPR}.

\begin{backmatter}
\bmsection{Funding} Shanghai Pilot Program for Basic Research (TQ20220104); National Natural Science Foundation of China (62175064, 62235019, 62035005); Natural Science Foundation of Chongqing (CSTB2023NSCQ-JQX0011, CSTB2022TIAD-DEX0036); Shanghai Municipal Science and Technology Major Project (2019SHZDZX01); Fundamental Research Funds for the Central Universities.

\bmsection{Disclosures} The authors declare no competing interests.

\bmsection{Data availability} The data that support the findings of this study are available from the corresponding author upon reasonable request.

\end{backmatter}

%%%%%%%%%%%%%%%%%%%%%%%%%%%%%%%%%%%%%%%%%%%%%%%%%%%%%%%%%%%%%%%

\end{document}